\documentclass[prd,twocolumn,nopacs,floatfix,amsmath,nofootinbib,amssymb,floatfix]{revtex4}
\usepackage{graphicx,color,dcolumn,booktabs,bm}
\usepackage{longtable,lscape}
\usepackage{txfonts}
\usepackage{overpic}
\usepackage{amssymb}
\usepackage{indentfirst}
\usepackage{feynmf}   %{feynmp}
\usepackage{slashed}  %for Feynman symbols
\usepackage{cases}
\usepackage{color}
\usepackage{multirow}
\usepackage{epstopdf}
\usepackage{graphicx,color,dcolumn,booktabs,bm}
\usepackage[colorlinks, citecolor=blue,anchorcolor=red,menucolor=red, linkcolor=red,filecolor=red,runcolor=red,urlcolor=blue,frenchlinks=red]{hyperref}

\graphicspath{{Figures/}} %

\begin{document}
%\begin{CJK}{GBK}{}

\title{Discovering an unquenched dynamics mechanism for charmonium scattering}
\author{Qi Huang$^{1,2}$}\email{06289@njnu.edu.cn}
\author{Rui Chen$^{3,2}$}\email{chenrui@hunnu.edu.cn}
\author{Jun He$^{1,2}$}\email{junhe@njnu.edu.cn}
\author{Xiang Liu$^{2,4,5,6,7}$}\email{xiangliu@lzu.edu.cn}
\affiliation{$^1$School of Physics and Technology, Nanjing Normal University, Nanjing 210023, China\\
$^2$Lanzhou Center for Theoretical Physics, Key Laboratory of Theoretical Physics of Gansu Province, Lanzhou University, Lanzhou 730000, China\\
$^3$Key Laboratory of Low-Dimensional Quantum Structures and Quantum Control of Ministry of Education, Department of Physics and Synergetic Innovation Center for Quantum Effects and Applications, Hunan Normal University, Changsha 410081, China\\
$^4$School of Physical Science and Technology, Lanzhou University, Lanzhou 730000, China\\
$^5$Key Laboratory of Quantum Theory and Applications of MoE, Lanzhou University, Lanzhou 730000, China\\
$^6$MoE Frontiers Science Center for Rare Isotopes, Lanzhou University, Lanzhou 730000, China\\
$^7$Research Center for Hadron and CSR Physics, Lanzhou University and Institute of Modern Physics of CAS, Lanzhou 730000, China}

\begin{abstract}

In this work, we propose a new mechanism for charmonium scattering that utilizes the internal structure of charmonium. By capturing light flavor quarks and anti-quarks from the vacuum, charm and anti-charm quarks form virtual charmed mesons, which mediate an effective one-boson exchange process. This approach accurately reproduces the di-$J/\psi$ invariant mass spectrum observed by CMS and LHCb, demonstrating its validity. Our mechanism offers a comprehensive framework for understanding charmonium scattering and is applicable to the scattering problems involving all fully heavy hadrons, an area of increasing interest. 
\end{abstract}

\maketitle

\section{Introduction}

Scattering problems, as a fundamental topic in physics, offering critical insights into the interactions and internal structures of matter. Using this approach, Yukawa predicted the pion meson \cite{Yukawa:1935xg}, which mediates the nuclear force, establishing a cornerstone of modern nuclear physics. In this century, advances in experimental techniques have led to the discovery of numerous new hadronic states, presenting unique opportunities to explore hadron scattering. Noteworthy achievements include the identification of exotic molecular states such as $P_c$ states \cite{LHCb:2015yax,LHCb:2019kea} and 
$T_{cc}(3875)^+$ ~\cite{LHCb:2021vvq,LHCb:2021auc}. These discoveries not only mark the advent of 'Particle 2.0' but also enhance our understanding of the nonperturbative aspects of the strong force \cite{Chen:2016qju,Richard:2016eis,Chen:2016spr,Lebed:2016hpi,Esposito:2016noz,Guo:2017jvc,Ali:2017jda,Brambilla:2019esw,Liu:2019zoy,Chen:2022asf}.

Recently, the CMS Collaboration reported a high-precision measurement of the invariant mass spectrum of di-$J/\psi$ \cite{CMS:2023owd}, confirming the previous observation of $X(6900)$
by the LHCb Collaboration \cite{LHCb:2020bwg} and revealing additional enhancements in the di-$J/\psi$ invariant mass spectrum. Furthermore, ATLAS has joined this effort, alongside LHCb and CMS \cite{ATLAS:2023bft}. This new finding surprised the community, prompting various explanations, including exotic hadronic states \cite{liu:2020eha,Tiwari:2021tmz,Lu:2020cns,Faustov:2020qfm,Zhang:2020xtb,Li:2021ygk,Liu:2021rtn,Giron:2020wpx,Karliner:2020dta,Wang:2020ols,Ke:2021iyh,Zhu:2020xni,Jin:2020jfc,Yang:2021hrb,Albuquerque:2020hio,Albuquerque:2021erv,Wu:2022qwd,Asadi:2021ids,Yang:2020wkh,Chen:2020xwe,Deng:2020iqw,Gordillo:2020sgc,Zhao:2020zjh,Faustov:2021hjs,Wang:2021kfv}
and coupled-channel dynamics generation \cite{Wang:2020wrp,Wang:2020tpt,Wang:2022jmb,Gong:2020bmg,Dong:2020nwy,Dong:2021lkh,Liang:2021fzr,Lu:2023aal,Guo:2020pvt,Zhuang:2021pci}. However, a substantial issue—charmonium scattering—has not been fundamentally resolved, despite some tentative efforts \cite{Wang:2020wrp,Wang:2020tpt,Wang:2022jmb,Gong:2020bmg,Dong:2020nwy,Dong:2021lkh,Liang:2021fzr,Lu:2023aal,Guo:2020pvt,Zhuang:2021pci}.
The highly effective one-boson ($\pi$, $\sigma$, $\rho$, $\cdots$) exchange mechanism for describing hadron scattering fails in quantitatively characterizing charmonium scattering, as charmonium is a typical example of a fully heavy hadron. Thus, studying how charmonium undergoes scattering has become a particularly typical issue in current scattering problems and urgently requires a solution.

\begin{figure}[htbp]
    \centering    \includegraphics[width=1.0\linewidth]{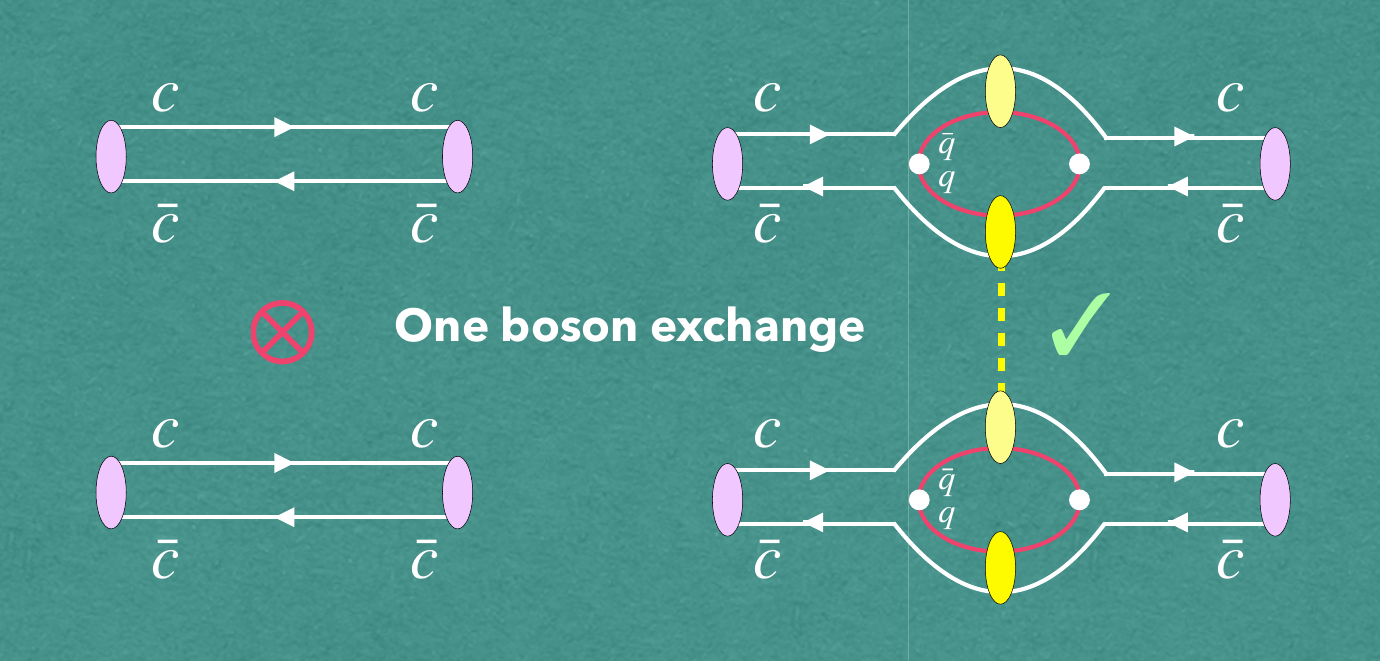}
    \caption{A new approach enabling one-boson exchange for effective charmonium scattering. The cross symbol on the left indicates that direct light flavor one-boson exchange ($\pi,\rho,\sigma,~\cdots$) cannot occur.}\label{fig:idea}
\end{figure}

In this work, we propose a new mechanism for charmonium scattering. As illustrated in Fig.~\ref{fig:idea}, the charm and anti-charm quarks within charmonium can capture light flavor anti-quark and quark, respectively, created from the vacuum to form virtual charmed mesons. These virtual mesons provide a source that enables one-boson exchange to be effective for charmonium scattering. A direct application of this new mechanism successfully reproduces the di-$J/\psi$ invariant mass spectrum data from CMS and LHCb, demonstrating the mechanism's validity. We should emphasize that this mechanism is not limited to charmonium scattering; it can also be applied to investigate the scattering problems involving all fully heavy hadrons, a topic that is gaining attention from the community, including lattice research groups \cite{Lyu:2021qsh,Mathur:2022ovu}.

\section{Application to reproduce the CMS and LHCb data}

The di-$J/\psi$ invariant mass spectrum observed by the CMS and LHCb Collaborations exhibit significant features around the dip near 6.8 GeV and the peak near 6.9 GeV, corresponding closely to the $J/\psi \psi(3686)$ and $J/\psi \psi(3770)$ thresholds, respectively. In contrast, the bumps near 6.6 GeV and 7.3 GeV do not align with any thresholds involving particles from the $\psi$ family. Therefore, to capture the main features, we consider couplings among three channels: ${J/\psi J/\psi, J/\psi \psi(3686), J/\psi \psi(3770)}$. As for the di-$\eta_c$ and di-$\chi_{cJ}$ channels, although there may exist thresholds that close to the observed peaks, due to their much lower production cross sections \cite{Bodwin:1994jh,ATLAS:2015zdw,ATLAS:2014ala,CMS:2012qwg,LHCb:2013ofo,He:2024lrb}, we neglect their effects in this work.

\begin{figure}[htbp]
    \centering
    \includegraphics[width=1.0\linewidth]{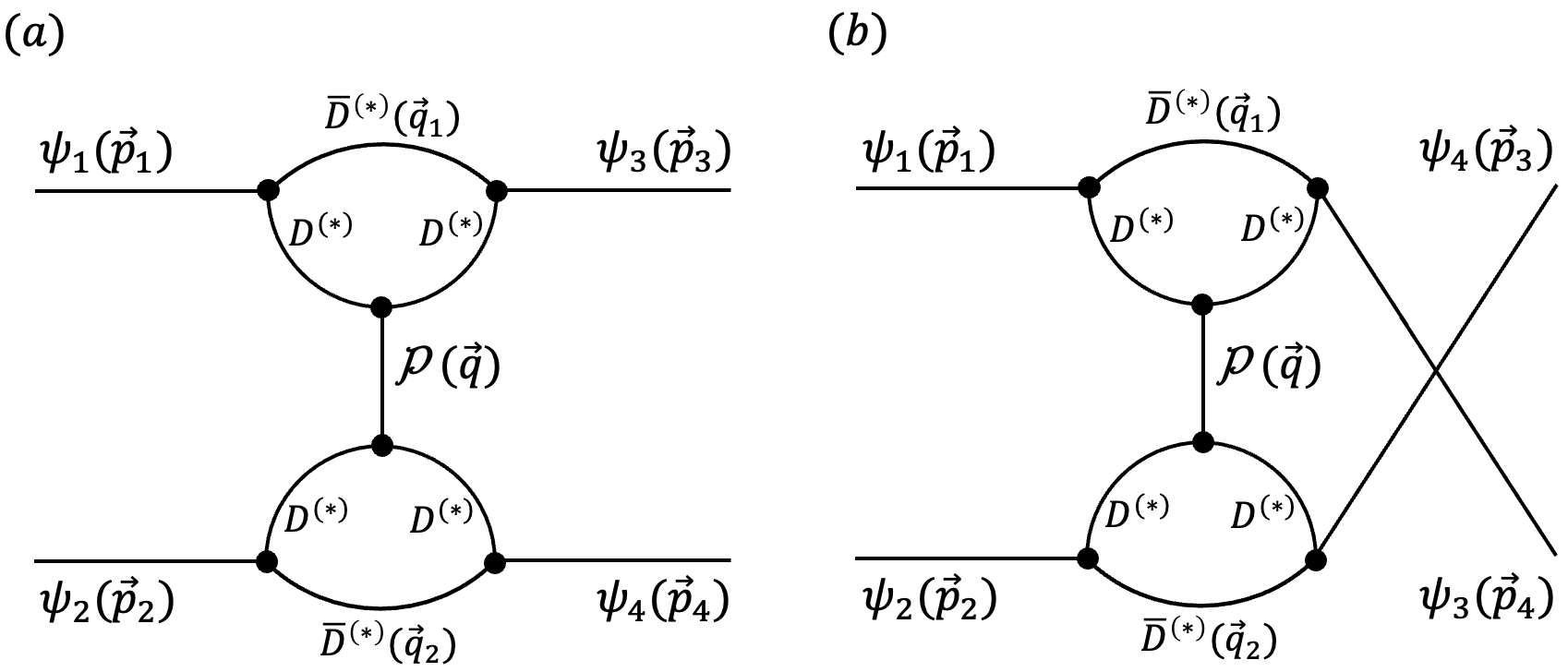}
    \caption{The scattering mechanisms of the $\psi_1 \psi_2 \to \psi_3 \psi_4$ process involve $D$ meson loops, with diagram $(a)$ representing the $t$-channel contribution and diagram $(b)$ representing the $u$-channel contribution. Here, $\psi_i$ can be $J/\psi$, $\psi(3686)$, or $\psi(3770)$, and $\mathcal{P}$ can be $\eta$ or $\sigma$. Furthermore, when $\psi_1 = \psi_2$ and $\psi_3 = \psi_4$, there is no $u$-channel contribution.}\label{fig:mechanism}
\end{figure}

When one-boson exchange is introduced at the quark level, as depicted in Fig.~\ref{fig:idea}, the scattering mechanism between two charmonia can proceed as shown in Fig.~\ref{fig:mechanism}. Initially, the two charmonia, $\psi_1$ and $\psi_2$, couple into pairs of virtual $D$ mesons. Within each pair, one $D$ meson acts as a spectator, while the other engages in an interaction mediated by the exchange of a light hadron, $\mathcal{P}$. Subsequently, the interacting virtual $D$ mesons combine with the spectators to form the final charmonia, $\psi_3$ and $\psi_4$ \iffalse\textcolor{red}{\footnote{\textcolor{red}{There exists a possibly more direct box loop diagram, where the four externel charmonia are linked by four internel $D$-mesons. However, according to the classical Yukawa theory \cite{Yukawa:1935xg}, this diagram has only about 0.1 fm effective range, which is much smaller than the characteristic size of a molecular state. Thus, we will not consider this mechanism here.}}}.\fi

As shown in Fig.~\ref{fig:mechanism}, we have neglected the contributions from charmed-strange meson loops. This is primarily due to the significantly lower creation rate of $s\bar{s}$ quark pairs compared to $u\bar{u}$ or $d\bar{d}$ quark pairs, with a ratio of approximately ${1}/{\sqrt{3}}$. Additionally, the higher threshold for producing charmed-strange meson pairs leads to further suppression on loop function. Therefore, to focus on the primary physics, we exclude contributions from charmed-strange meson loops.

Furthermore, Fig.~\ref{fig:mechanism} illustrates that the exchanged light hadron, $\mathcal{P}$, is limited to $\sigma$ and $\eta$. This restriction arises because the final interaction between two charmonia and a light hadron still must comply with the conservation laws of the strong interaction. Given that the mass of the exchanged particle should be as small as possible, only $\sigma$ and $\eta$ are viable candidates, they can contribute the intermediate and long range interactions, whereas, the direct charmed mesons exchanges provide a very short range interactions due to their large masses \cite{Yukawa:1935xg}, although it seems to be more efficient interactions to study the charmonia interactions. Our result demonstrates
that within our model the sigma-exchange dominates over the eta-exchange, as
expected\footnote{With the fit parameters listed in Table~\ref{tab:fit}, we calculate the ratio of the $|\mathcal{M}^J_{J/\psi J/\psi}(\sqrt{s})|^2$, defined in Eq.(\ref{eq:amplitude}), that purely caused by $\sigma$ and $\eta$ exchanges. The numerical showed that for both $0^{++}$ and $2^{++}$ cases, when $\sqrt{s}$ varies from 6.3 to 9.0 GeV, this ratio is always much larger than one.}.

We perform the entire calculation via the effective Lagrangian approach, with relevant Lagrangians being \cite{Cheng:1992xi,Yan:1992gz,Wise:1992hn,Burdman:1992gh,Casalbuoni:1996pg}
\iffalse
\begin{eqnarray}
    \mathcal{L}_{D^{(\ast)} D^{(\ast)} \eta} &=& -\frac{2 g}{\sqrt{6} f_\pi} \left(i \varepsilon_{\alpha \mu \nu \lambda} v^\alpha D^{\ast \mu} D_a^{\ast \lambda} \partial^\nu \eta + D D_{\lambda}^{*} \partial^\lambda \eta \right),\\
    \mathcal{L}_{\bar{D}^{(\ast)} \bar{D}^{(\ast)} \eta} &=& \frac{2 g}{\sqrt{6} f_\pi} \left( i\varepsilon_{\alpha \mu \nu \lambda} v^\alpha \bar{D}^{\ast \mu} \bar{D}^{\ast \lambda} \partial^\nu \eta + \bar{D}_{\lambda}^{\ast} \bar{D} \partial^\lambda \eta\right), \\
    \mathcal{L}_{D^{(\ast)} D^{(\ast)} \sigma} &=& -2 g_s D D \sigma + 2 g_s D^\ast \cdot D^\ast \sigma, \\
    \mathcal{L}_{\bar{D}^{(\ast)} \bar{D}^{(\ast)} \sigma} &=& -2 g_s \bar{D} \bar{D} \sigma + 2 g_s \bar{D}^\ast \cdot \bar{D}^\ast \sigma, \\
    \mathcal{L}_{D^\ast \bar{D}^\ast J / \psi} &=& -i g_{D^\ast D^\ast \psi}\left[\psi \cdot \bar{D}^\ast \overleftrightarrow{\partial} \cdot D^\ast\right. \nonumber\\
    && \left.\left.-\psi^\mu \bar{D}^\ast \cdot \overleftrightarrow{\partial}{ }^\mu D^\ast+\psi^\mu \bar{D}^\ast \cdot \overleftrightarrow{\partial} D^{\ast \mu}\right)\right], \\
    \mathcal{L}_{D^\ast \bar{D} J / \psi} &=& g_{D^\ast D \psi} \epsilon_{\beta \mu \alpha \tau} \partial^\beta \psi^\mu\left(\bar{D} \overleftrightarrow{\partial}{ }^\tau D^{\ast \alpha}+\bar{D}^{\ast \alpha} \overleftrightarrow{\partial^\tau} D\right), \\
    \mathcal{L}_{D \bar{D} J / \psi} &=& i g_{D D} \psi \cdot \bar{D} \overleftrightarrow{\partial} D.
\end{eqnarray}
\fi
\begin{eqnarray}
    \mathcal{L}_{D^{(\ast)} D^{(\ast)} \eta} &=& -\frac{2 g}{\sqrt{6} f_\pi} \left(i \varepsilon_{\alpha \mu \nu \lambda} v^\alpha D^{\ast \mu} D_a^{\ast \lambda} \partial^\nu \eta + D D_{\lambda}^{*} \partial^\lambda \eta \right),\\
    \mathcal{L}_{D^{(\ast)} D^{(\ast)} \sigma} &=& -2 g_s D D \sigma + 2 g_s D^\ast \cdot D^\ast \sigma, \\
    \mathcal{L}_{D^\ast \bar{D}^\ast J / \psi} &=& -i g_{D^\ast D^\ast \psi}\left[\psi \cdot \bar{D}^\ast \overleftrightarrow{\partial} \cdot D^\ast-\psi^\mu \bar{D}^\ast \cdot \overleftrightarrow{\partial}{ }^\mu D^\ast\right. \nonumber\\
    && \left.\left.+\psi^\mu \bar{D}^\ast \cdot \overleftrightarrow{\partial} D^{\ast \mu}\right)\right], \\
    \mathcal{L}_{D^\ast \bar{D} J / \psi} &=& g_{D^\ast D \psi} \epsilon_{\beta \mu \alpha \tau} \partial^\beta \psi^\mu\left(\bar{D} \overleftrightarrow{\partial}{ }^\tau D^{\ast \alpha}+\bar{D}^{\ast \alpha} \overleftrightarrow{\partial^\tau} D\right), \\
    \mathcal{L}_{D \bar{D} J / \psi} &=& i g_{D D} \psi \cdot \bar{D} \overleftrightarrow{\partial} D.
\end{eqnarray}
Here, $v$ is the velocity of heavy quark. $g = 0.59$, $f_\pi = 132$ MeV, and $g_s = 0.76$ are the fixed coupling constants~\cite{Ding:2021igr,Falk:1992cx,Isola:2003fh,Liu:2009qhy,Chen:2019asm}. $g_{D^\ast D^\ast \psi}$, $g_{D D \psi}$, and $g_{D^\ast D \psi}$ can be related to one single parameter $f_\psi$ as $\frac{g_{D^\ast D^\ast \psi}}{m_{D^\ast}} = \frac{g_{D D \psi}}{m_{D}} = g_{D^\ast D \psi} = \frac{m_\psi}{m_D f_\psi}$ due to heavy quark effective theory.

The amplitudes for the diagrams shown in Fig.~\ref{fig:mechanism} can be generally expressed as
\begin{eqnarray}
    \mathcal{M}_{(a)} &=& 16 \epsilon_1^\mu \epsilon_2^\nu \epsilon_3^{\ast\alpha} \epsilon_4^{\ast\beta} \frac{1}{q^2-m_\mathcal{P}^2+i m_\mathcal{P} \Gamma_\mathcal{P}} \mathcal{F}^2_q(q) \nonumber\\
    &&\times \int \frac{d^4 q_1}{(2\pi)^4} \frac{\mathcal{V}_{\mu\alpha}}{\prod_i (p_i^2-m_i^2+im_i\Gamma_i)} \mathcal{F}_{q_1}(q_1) \nonumber\\
    &&\times \int \frac{d^4 q_2}{(2\pi)^4} \frac{\mathcal{V}_{\nu\beta}}{\prod_j (p_j^2-m_j^2+im_j\Gamma_j)} \mathcal{F}_{q_2}(q_2),\label{eq:Mt} 
\end{eqnarray}
\begin{eqnarray}
    \mathcal{M}_{(b)} &=& 16 \epsilon_1^\mu \epsilon_2^\nu \epsilon_4^{\ast\alpha} \epsilon_3^{\ast\beta} \frac{1}{q^2-m_\mathcal{P}^2+i m_\mathcal{P} \Gamma_\mathcal{P}} \mathcal{F}_q^2(q) \nonumber\\
    &&\times \int \frac{d^4 q_1}{(2\pi)^4} \frac{\mathcal{V}_{\mu\alpha}}{\prod_i (p_i^2-m_i^2+im_i\Gamma_i)} \mathcal{F}_{q_1}(q_1) \nonumber\\
    &&\times \int \frac{d^4 q_2}{(2\pi)^4} \frac{\mathcal{V}_{\nu\beta}}{\prod_j (p_j^2-m_j^2+im_j\Gamma_j)} \mathcal{F}_{q_2}(q_2),\label{eq:Mu}
\end{eqnarray}
where $\mathcal{V}_{\gamma\delta}$ denotes the Lorentz structure of the loop integral, $\mathcal{F}_{q}(q)$ is the form factor for the exchanged light hadron $\mathcal{P}$ to account for the off-shell effect, and $\mathcal{F}_{q_i}(q_i)$ is a form factor used to avoid ultraviolet divergence in the loop integral. We use the widely adopted monopole form for $\mathcal{F}_q(q)$ as $\mathcal{F}_q(q) = \frac{\Lambda_q^2 - m_q^2}{\Lambda_q^2 - q^2}$, where $\Lambda_q$ is parametrized as $\Lambda_q = m_q + \alpha_q \Lambda_{\mathrm{QCD}}$, with $\Lambda_{\mathrm{QCD}} = 220$ MeV and $\alpha_q$ is an adjustable parameter. For $\mathcal{F}_{q_i}(q_i)$, we adopt $\mathcal{F}_{q_i}(q_i) = \exp(-\vec{q}_i^{~2}/\Lambda_i^2)$.

For both Eq.~(\ref{eq:Mt}) and Eq.~(\ref{eq:Mu}), we multiply them with an overall factor of 16. It is because each $D$ meson loop in Fig.~\ref{fig:mechanism} can involve not only $D^{(\ast)} D^{(\ast)} \bar{D}^{(\ast)}$, but also $\bar{D}^{(\ast)} \bar{D}^{(\ast)} D^{(\ast)}$ configurations. Furthermore, for both of these cases, there have charged and neutral situations. Thus, each loop has a total of 4 configurations, resulting in 16 configurations for each diagram in Fig.~\ref{fig:mechanism}. We want to emphasize here that for $J/\psi J/\psi$ interactions in this model, although it seems the diagram may be suppressed by the masses of the exchanged mesons, like $1/(m_{J/\psi}^2 - 4 m_D^2) \times 1/m_\mathcal{P}^2$, this suppression can be overcame by the summations on the amplitudes with all the possible intermediate $D^{(\ast)}$ and $\bar{D}^{(\ast)}$ combinations in addition with a large but not unreasonable cutoff in form factor. In particular, when the discussed charmonia has a larger mass, the two $D$-mesons connected with it will be more close to their mass shells, which can reduce the suppression factor significantly.

In this work, we consider only the $S$-wave behavior in the di-$J/\psi$ spectrum. Therefore, we first perform a partial wave expansion on the amplitude to get the interaction kernel, which can be calculated as
\begin{eqnarray}
    \mathcal{K}^{J}_{\lambda,\lambda^\prime}(\sqrt{s},k,k^\prime) = 2\pi \int d \cos\theta d_{\lambda \lambda^\prime}^J (\theta) \mathcal{M}_{\lambda,\lambda^\prime}(\sqrt{s},\vec{k},\vec{k}^\prime)
\end{eqnarray}
with $k^{(\prime)} = |\vec{k}^{~(\prime)}|$ being the module of the three-momenta of initial and final particles in the center-of-mass system, $\lambda$ and $\lambda^\prime$ is the helicity indices of the initial and final states, respectively, and $d_{\lambda,\lambda^\prime}^J(\theta)$ is the Wigner-$d$ matrix with $\theta$ being the relative angle between $\vec{k}$ and $\vec{k}'$.

Next, we substitute the partial wave interaction kernel into the coupled-channel Lippmann-Schwinger equation
\begin{eqnarray}
    T^J_{\lambda,\lambda^\prime}(\sqrt{s},k,k^\prime) &=& \mathcal{K}^J_{\lambda,\lambda^\prime}(\sqrt{s},k,k^\prime) + \sum_{\lambda^{\prime\prime}} \int \frac{k^{\prime\prime 2}d k^{\prime\prime}}{(2\pi)^3}\nonumber\\
    &&\times \mathcal{K}^J_{\lambda,\lambda^{\prime\prime}}(\sqrt{s},k,k^{\prime\prime}) G_0(\sqrt{s},k^{\prime\prime})\nonumber\\
    &&\times T^J_{\lambda^{\prime\prime},\lambda^\prime}(\sqrt{s},k^{\prime\prime},k^\prime),
\end{eqnarray}
then, with the extracted on-shell coupled-channel $T$-matrices $T^J(\sqrt{s})$, the transition amplitude can be written as \cite{Gong:2020bmg,Dong:2020nwy}
\begin{eqnarray}
    \mathcal{M}^J_{J/\psi J/\psi}(\sqrt{s}) = \alpha e^{-\beta s} \left( 1 + \sum_i r_i e^{i \phi_i} G_{0i}(\sqrt{s})T_{i,J/\psi J/\psi}^J(\sqrt{s}) \right),\label{eq:amplitude}
\nonumber
\end{eqnarray}
where $G_0$ is the free two particle Green's function, $G_{0i}$ means the free Green's function of $i$-th channel, $\alpha$ represents an overall factor, $r_i$ and $\phi_i$ describe the relative production strength and phase angle for channel $i$, respectively, and $\beta$ is used to simulate the background. Although the introduction of $r_i$ and $\phi_i$ will violate the unitary, however, unitarity is strictly maintained only within the scattering sector---calculated via the Lippmann--Schwinger equation for the coupled channels \(\{J/\psi J/\psi, J/\psi \psi(3686), J/\psi \psi(3770)\}\). When this unitary scattering amplitude is combined with the parameterized, non-calculated production amplitude to form the full observable amplitude, overall unitarity is broken, as we do not perform a fully coupled-channel calculation including the \( p p \) state.

Finally, the di-$J/\psi$ spectrum can be calculated as \cite{Gong:2020bmg,Dong:2020nwy}
\begin{eqnarray}
    \sigma(s) = \frac{\left|\vec{p}_{J/\psi}\right|}{8 \pi s} \left|\mathcal{M}^J_{J/\psi J/\psi}(\sqrt{s}) \right|^2,
\end{eqnarray}
with $\vec{p}_{J/\psi}$ being the three momentum of final $J/\psi$ in the center-of-mass frame.

We primarily focus on the CMS data, utilizing the LHCb data as supplementary due to the higher precision of the CMS muon detector. The fit parameters are as follows, excluding the semi-fixed overall factors $\alpha$ and $\beta$ which control the lineshape and magnitude of background, for each channel $i$, there are two parameters: the phase angle $\phi_i$ and the relative production strength $r_i$.

Our model includes cut-off dependent form factors, making the scattering amplitudes inherently cut-off dependent. To avoid lengthy fitting processes, we fix all cut-offs ($\alpha_q$ and $\Lambda_i$) to one and treat the decay constants $f_{\psi}$ as free parameters. This approach absorbs the cut-off dependence into the coupling constants, adhering to the principles of renormalization. Consequently, there are three additional parameters: $f_{J/\psi}$, $f_{\psi(3686)}$, and $f_{\psi(3770)}$, resulting in a total of 11 parameters for our fit.

\begin{figure}[htbp]
    \centering
    \includegraphics[width=0.98\linewidth]{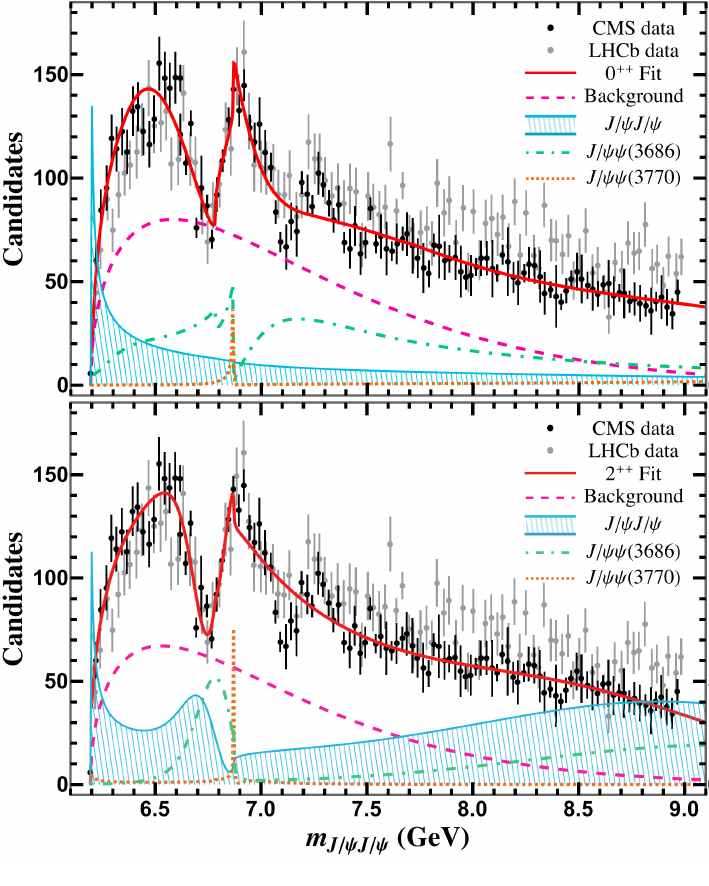}
    \caption{The fit results on experimental data are presented, with the top panel for $0^{++}$ configuration and the bottom panel for $2^{++}$ configuration. In each panel, the dark and light grey error bars represent the CMS data and LHCb data respectively, and the red solid line indicates the fit result. The magenta dashed line, light blue line filled with shadows, light green dot-dashed line, and orange dotted line show the contributions of background, $J/\psi J/\psi$ channel, $J/\psi \psi(3686)$ channel, and $J/\psi \psi(3770)$ channel, respectively.}
  \label{fig:fit}
\end{figure}

For the di-$J/\psi$ invaraint mass spectrum, considering only the $S$-wave effective potential in the scattering, there are two possible quantum numbers: $0^{++}$ and $2^{++}$. The $1^{++}$ configuration is excluded because, although $1 \otimes 1 \rightarrow 1$ is possible, it cannot form a state with $C=+$ parity under $S$-wave conditions. Table~\ref{tab:fit} presents the results of the two sets of fit parameters for the $0^{++}$ and $2^{++}$ configurations. The corresponding fit results \and extracted pole positions, based on the parameters in Tab.~\ref{tab:fit}, are shown in Fig.~\ref{fig:fit} and Tab.~\ref{tab:pole}, respectively.

\begin{table}[htpb]
    \renewcommand\arraystretch{1.5}
    \centering
    \caption{Fit parameters for reproducing the CMS experimental data in the $0^{++}$ and $2^{++}$ configurations. Parameters marked with an asterisk ($\ast$) are fixed during the fit.}\label{tab:fit}
    \begin{tabular}{ccccc}
    \toprule[1pt]
    Parameter &$~$& $0^{++}$ &$~$& $2^{++}$\\
    \midrule[1pt]
    $\alpha$ &$~$& 617.338$^\ast$ &$~$& 981.439$^\ast$\\
    $\beta$ &$~$& 0.0447$^\ast$ &$~$& 0.0528$^\ast$\\
    $r_{J/\psi J/\psi}$ &$~$& 1.101 &$~$& 0.429\\
    $\phi_{J/\psi J/\psi}$ &$~$& 2.979 &$~$& 3.179\\
    $r_{J/\psi \psi(3686)}$ &$\qquad$& 0.724 &$\qquad$& 0.487 \\
    $\phi_{J/\psi \psi(3686)}$ &$\qquad$& 2.484 &$\qquad$& 2.745\\
    $r_{J/\psi \psi(3770)}$ &$\qquad$& 1.508 &$\qquad$& 0.268\\
    $\phi_{J/\psi \psi(3770)}$ &$\qquad$& 0.196 &$\qquad$& 0.542\\
    $f_{J/\psi}$ & $\qquad$ & 0.378 GeV & $\qquad$ & 0.428 GeV\\
    $f_{\psi(3686)}$ & $\qquad$ & 0.237 GeV & $\qquad$ & 0.116 GeV\\
    $f_{\psi(3770)}$ & $\qquad$ & 0.760 GeV & $\qquad$ & 0.267 GeV\\
    \midrule[1pt]
    $\chi^2/d.o.f$ & $\qquad$ & 2.067 & $\qquad$ & 2.268 \\
    \bottomrule[1pt]
    \end{tabular}
\end{table}

\iffalse

\begin{table*}[htbp]
    \renewcommand\arraystretch{1.5}
    \caption{\textcolor{magenta}{The pole positions extracted from our fit parameters given in Tab.~\ref{tab:fit}, where the bold one without error and with same value in both $0^{++}$ and $2^{++}$ cases comes from triangle singularity. Here, all the values are presented in unit GeV.}}
    \label{tab:pole}
    \centering 
    \scalebox{1.0}{
    \begin{tabular}{c|ccccc}
        \hline \hline
        $0^{++}$ case & $(6.204 \pm 0.024) - (0.023 \pm 0.023)i$ & $(6.789 \pm 0.017) - (0.051 \pm 0.051)i$ & $(6.871 \pm 0.011) - (0.025 \pm 0.025)i$ & $\boldsymbol{6.95}$ \\
        \hline
        $2^{++}$ case & $(6.198 \pm 0.007) - (0.007 \pm 0.007)i$ & $(6.783 \pm 0.009) - (0.008 \pm 0.008)i$ & $(6.871 \pm 0.010) - (0.038 \pm 0.038)i$ & $\boldsymbol{6.95}$ \\
        \hline
        \hline
    \end{tabular}}
\end{table*}
\fi

\renewcommand\tabcolsep{0.5cm}
\begin{table}[htbp]
    \renewcommand\arraystretch{1.5}
    \caption{The pole positions $m-\frac{\Gamma}{2}i$ extracted from our fit parameters given in Table~\ref{tab:fit} for the $0^{++}$ and $2^{++}$ cases. All values are given in units of MeV. Here, the uncertainties in the pole positions come from the variations of the cutoffs within 50 MeV.}
    \label{tab:pole}
    \centering 
    \scalebox{1.0}{
    \begin{tabular}{c|c}
        \toprule[1pt]
        $0^{++}$ case & $2^{++}$ case\\\midrule[1pt]
        $(6204 \pm 24) - (23 \pm 23)i$   
            &$(6198 \pm 7) - (7 \pm 7)i$\\
        $(6789 \pm 17) - (51 \pm 51)i$ 
            & $(6783 \pm 9) - (8 \pm 8)i$\\
        $(6871 \pm 11) - (25 \pm 25)i$ 
            & $(6871 \pm 10) - (38 \pm 38)i$\\
        \bottomrule[1pt]
    \end{tabular}}
\end{table}

As shown in Fig.~\ref{fig:fit}, our model successfully reproduces the overall features of the di-$J/\psi$ invariant mass spectrum. This indicates that our proposed scattering mechanism between two charmonium states is reasonable, providing a new approach to understanding hadron scattering, especially for fully heavy hadrons. Furthermore, both the lineshape of the black solid lines and the $\chi^2/d.o.f$ values in Table~\ref{tab:fit} suggest a slight preference for the $0^{++}$ configuration in the di-$J/\psi$ invariant mass spectrum, but $2^{++}$ configuration is also possible. Here, the $\chi^2/d.o.f$ values are not close to 1, primarily because there are no thresholds near 6.6 GeV and 7.3 GeV in our fit, leading to mismatches in these regions. Despite this, our primary goal of validating the new scattering mechanism is achieved through the correspondence with experimental results.

\iffalse
In addition, Fig.~\ref{fig:fit} reveals a distinct structure around 6.94~GeV, \textcolor{blue}{which is close to the $J/\psi\psi(3770)$ threshold. As illustrated in Fig.~\ref{fig:mechanism}, this phenomenon can be naturally attributed to the anomalous threshold generated by the $D\bar{D}$ triangle diagram involving $\psi(3770)$. Since $\psi(3770)$ lies above the $D\bar{D}$ threshold and couples strongly to this pair, the triangle loop amplitude develops an anomalous branch cut at $t = -**~\mathrm{GeV}^2$. When the center-of-mass energy exceeds $\sqrt{s} \approx 6.94~\mathrm{GeV}$, this anomalous singularity enters the integration region of the partial-wave projection, leading to a significant enhancement in the $J/\psi\psi(3770)$ scattering amplitude near the $X(6900)$ peak \cite{Lu:2023aal,Zhuang:2021pci}. Unlike a conventional threshold cusp, this enhancement stems from the dynamical propagation of the anomalous cut into the $s$-channel, and its precise position depends on the masses of the intermediate $D$ mesons. Although the triangle contribution alone may peak slightly above the observed $X(6900)$ mass, its interference with background amplitudes can shift the peak downward by several tens of MeV. Therefore, such an anomalous threshold effect provides a plausible mechanism for the formation of the $X(6900)$ structure and should be considered an essential feature of the molecular interpretation involving $\psi(3770)$. Future high-precision measurements around 6.9~GeV will be crucial for testing this scenario.}
\fi

In addition, Fig.~\ref{fig:fit} reveals a twist at 6.865 GeV, which is just located at the $J/\psi \psi(3770)$ threshold. Obviously, from the diagram given in Fig.~\ref{fig:mechanism}, it can be inferred that such phenomenon is attributed to the $\psi(3770)$, whose mass is above the $D\bar{D}$ threshold and thus can decay into a pair of real $D$ mesons. As a result, when the scattering energy reach up to the $J/\psi \psi(3770)$ threshold, rescattering between coupled channels exceed the branch point located at the $J/\psi \psi(3770)$ threshold, which can cause a cusp at this threshold~\cite{Wang:2020wrp}. In addition, since $\psi(3770)$ can be a real particle at this time, when the two charmed mesons linked with it are $D\bar{D}$, the branch point located at $D\bar{D}$ threshold is simultaneously triggered, which may result into further enhancements and generate a tall but very narrow peak just around the $J/\psi \psi(3770)$ threshold, and consequently only slightly modifies the shape of the peak near 6.9~GeV. As for the triangle singularity, based on the algorithm we used in our previous work~\cite{Huang:2020kxf}, it can occur when the intermediate states in the triangle loop is $D^\ast \bar{D} D$, with $D^\ast \bar{D}$ being linked with charmonia, and $D^\ast D$ connected to the light exchanged boson. However, at this time, the kinematical limitation needs the charmonia should be an intermediate virtual particle with energy in a range from 3.872 to 3.875 GeV, and the exchanged light boson should have energy less than 0.14 GeV. Apparently, such an effect is highly possible to be suppressed, since either the charmonia or the exchanged light boson is deviated from the center value of mass. Thus, \iffalse When $\psi_3 = J/\psi$ and the remaining intermediate meson in the triangle loop is also a $D$ meson, the anomalous threshold effect occurs, resulting in the twist around 6.9 GeV. \textcolor{magenta}{We emphasize that, since the widths of the $D$ mesons in the loop are nearly zero, this anomalous-threshold effect produces a tall but very narrow peak, and consequently only slightly modifies the shape of the peak near 6.9~GeV}. Actually, this kind of threshold effect has been extensively discussed in previous studies \cite{Lu:2023aal,Zhuang:2021pci}, and our results are consistent with Ref. \cite{Lu:2023aal}.\fi the presence of such a strong threshold effect is a crucial feature of our model. More precise future experimental measurements around 6.9 GeV in the future would greatly aid in validating our model.

\textcolor{magenta}{}

Moreover, the wide peak around 6.6 GeV can be reproduced by interference effects without introducing additional channels. The sharp resonance-like structure in $\sigma_{J/\psi J/\psi}$ in both configurations suggests a potential $J/\psi J/\psi$ bound state, as it is not reflected in the overall spectrum due to phase space limitations. This conclusion aligns with Refs. \cite{Dong:2020nwy,Dong:2021lkh,Liang:2021fzr}. Importantly, if seeing Table~\ref{tab:pole} and Fig.~\ref{fig:fit} together, the dip near 6.8 GeV, particularly evident in the $J/\psi \psi(3686)$ contribution in the $2^{++}$ fit, indicates the existence of a state similar to $\Upsilon(10753)$, which also shows as a dip in $e^+e^- \to b\bar{b}$ cross section \cite{Dong:2020tdw}. This pole around 6.8 GeV is also identified in Refs. \cite{Dong:2020nwy,Liang:2021fzr,Guo:2020pvt,Zhuang:2021pci}. Thus, detailed future experimental analyses of this dip could further test our model.

Overall, it indicates that for either $0^{++}$ or $2^{++}$ configurations, there always exists a states below the $J/\psi J/\psi$ threshold. In addition, $X(6600)$ is not a real physical state but an effect of interference, while the dip around 6.8 GeV correspond to a pole mainly generated by the $J/\psi\psi(3686)$ channel. The peak around 6.9~GeV is dominantly generated by a physical state emerging from the $J/\psi\psi(3770)$ channel, but the strong threshold cusp effect there will also make an explicit twist on the shape of the peak. Due to the reason that our mechanism integrates nearly all the conclusions from previous studies into a unified framework, this initial success in understanding the scattering mechanism between two charmonium states motivates further study and broader application of our model. This work represents the first step in that direction.

\section{Summary}

Recently, the CMS Collaboration reported a high-precision measurement of the di-$J/\psi$ invariant mass spectrum \cite{CMS:2023owd}, which confirmed the previous observation of $X(6900)$ by the LHCb Collaboration \cite{LHCb:2020bwg} and revealed additional enhancements. The ATLAS Collaboration has also contributed to this effort, reinforcing these findings \cite{ATLAS:2023bft}. These observations have spurred a variety of theoretical interpretations, including the existence of exotic hadronic states and dynamics generated by coupled channels \cite{liu:2020eha,Tiwari:2021tmz,Lu:2020cns,Faustov:2020qfm,Zhang:2020xtb,Li:2021ygk,Liu:2021rtn,Giron:2020wpx,Karliner:2020dta,Wang:2020ols,Ke:2021iyh,Zhu:2020xni,Jin:2020jfc,Yang:2021hrb,Albuquerque:2020hio,Albuquerque:2021erv,Wu:2022qwd,Asadi:2021ids,Yang:2020wkh,Chen:2020xwe,Deng:2020iqw,Gordillo:2020sgc,Zhao:2020zjh,Faustov:2021hjs,Wang:2021kfv}. However, a critical issue remains unresolved: the nature of charmonium scattering. Traditional one-boson exchange mechanisms, highly effective for describing hadron scattering, fall short in quantitatively characterizing charmonium scattering. This shortcoming is particularly pronounced in fully heavy hadron systems.

To solve this puzzle, in this work, we introduce a new mechanism for charmonium scattering. Our approach leverages the internal structure of charmonium, where charm and anti-charm quarks can capture light flavor anti-quarks and quarks from the vacuum, forming virtual charmed mesons. These virtual mesons act as mediators, enabling an effective one-boson exchange process for charmonium scattering. 

Our proposed mechanism successfully reproduces the di-$J/\psi$ invariant mass spectrum data from CMS and LHCb \cite{CMS:2023owd,LHCb:2020bwg}, validating its effectiveness. A significant advantage of this approach is its broad applicability. It is not limited to charmonium scattering but can be extended to investigate the scattering of all fully heavy hadrons, an area of increasing interest within the physics community, including lattice QCD research groups \cite{Lyu:2021qsh,Mathur:2022ovu}. With our new mechanism, we can easily predict that there should also exist several states in the di-$\chi_{cJ}$ system, which are all very closed to their corresponding thresholds. Furthermore, due to the heavy quark symmetry, our mechanism should also exist in $\Upsilon\Upsilon$, $J/\psi\Upsilon$, $\chi_{bJ}\chi_{bJ'}$, $\chi_{cJ}\chi_{bJ'}$ channels, which can result into the generations of near threshold states. Nevertheless, this approach addresses the limitations of traditional one-boson exchange mechanisms, which are inadequate for fully heavy hadron systems due to the absence of light quark exchanges. It lays the groundwork for a deeper understanding of scattering mechanisms in fully heavy hadron systems, with potential implications for broader applications in particle physics. Thus, We emphasize the importance of further experimental investigations, particularly high-precision measurements around key energy thresholds, to refine and validate our model.

\iffalse By incorporating virtual meson formation, our model provides a robust framework for understanding charmonium interactions.

This study provides a comprehensive framework for charmonium scattering, integrating well with existing experimental observations \cite{CMS:2023owd,LHCb:2020bwg} and theoretical efforts on scattering mechanisms \cite{Wang:2020wrp,Wang:2020tpt,Wang:2022jmb,Gong:2020bmg,Dong:2020nwy,Dong:2021lkh,Liang:2021fzr,Lu:2023aal,Guo:2020pvt,Zhuang:2021pci}. The success of our mechanism in explaining the di-$J/\psi$ spectrum sets a precedent for future studies on fully heavy hadron scattering. We emphasize the importance of further experimental investigations, particularly high-precision measurements around key energy thresholds, to refine and validate our model. Our work lays the groundwork for a deeper understanding of scattering mechanisms in fully heavy hadron systems, with potential implications for broader applications in particle physics.
\fi

\section{Note added}

We noticed that the CMS collaboration has measured the spin-parity of all-charm tetraquarks \cite{CMS:2025ecr}. The result showed that the quantum numbers of these states prefer $J^{PC}=2^{++}$, which corroborates the conclusions of this work.

\section*{Acknowledgements}This work is supported by the National Natural Science Foundation of China under Grant Nos. 12335001, 12247101, 12305139, 12305087, and 12475080, the National Key Research and Development Program of China under Contract No. 2020YFA0406400, the ‘111 Center’ under Grant No. B20063, the Natural Science Foundation of Gansu Province (No. 22JR5RA389, No. 25JRRA799), the fundamental Research Funds for the Central Universities (No. lzujbky-2023-stlt01), the project for top-notch innovative talents of Gansu province, Lanzhou City High-Level Talent Funding, and the Xiaoxiang Scholars Programme of Hunan Normal University.

%Qi Huang wants to thank Jia-Jun Wu, Hao-Jie Jing, Zheng-Li Wang, Chao-Wei Shen, Zhi-Feng Sun, Yu-Heng Wu, Shu-Yi Kong, Hong-Xia Huang and Jia-Lun Ping for very useful discussions. Qi Huang also wants to thank Ri-Qing Qian for useful technical support. 

\vfil


\begin{thebibliography}{}

    %\cite{Yukawa:1935xg}
    \bibitem{Yukawa:1935xg}
    H.~Yukawa,
    On the Interaction of Elementary Particles I, \href{https://academic.oup.com/ptps/article/doi/10.1143/PTPS.1.1/1878532}{Proc. Phys. Math. Soc. Jap. \textbf{17}, 48-57 (1935).}
    %doi:10.1143/PTPS.1.1
    %1261 citations counted in INSPIRE as of 22 Jul 2024

%\cite{LHCb:2015yax}
\bibitem{LHCb:2015yax}
R.~Aaij \textit{et al.} [LHCb],
Observation of $J/\psi p$ Resonances Consistent with Pentaquark States in $\Lambda_b^0 \to J/\psi K^- p$ Decays,
\href{https://journals.aps.org/prl/abstract/10.1103/PhysRevLett.115.072001}{Phys. Rev. Lett. \textbf{115}, 072001 (2015).}
%1778 citations counted in INSPIRE as of 22 Jul 2024

%\cite{LHCb:2019kea}
\bibitem{LHCb:2019kea}
R.~Aaij \textit{et al.} [LHCb],
Observation of a narrow pentaquark state, $P_c(4312)^+$, and of two-peak structure of the $P_c(4450)^+$,
\href{https://journals.aps.org/prl/abstract/10.1103/PhysRevLett.122.222001}{Phys. Rev. Lett. \textbf{122}, 222001 (2019).}


%\cite{LHCb:2021vvq}
\bibitem{LHCb:2021vvq}
R.~Aaij \textit{et al.} [LHCb],
Observation of an exotic narrow doubly charmed tetraquark,
\href{https://www.nature.com/articles/s41567-022-01614-y}{Nature Phys. \textbf{18}, 751-754 (2022).}

%\cite{LHCb:2021auc}
\bibitem{LHCb:2021auc}
R.~Aaij \textit{et al.} [LHCb],
Study of the doubly charmed tetraquark $T_{cc}^{+}$,
\href{https://www.nature.com/articles/s41467-022-30206-w}{Nature Commun. \textbf{13}, 3351 (2022).}

%\cite{Chen:2016qju}
\bibitem{Chen:2016qju}
H.~X.~Chen, W.~Chen, X.~Liu and S.~L.~Zhu,
The hidden-charm pentaquark and tetraquark states,
\href{https://doi.org/10.1016/j.physrep.2016.05.004}{Phys. Rept. \textbf{639}, 1-121 (2016).}

%\cite{Richard:2016eis}
\bibitem{Richard:2016eis}
J.~M.~Richard,
Exotic hadrons: review and perspectives,
\href{https://doi.org/10.1007/s00601-016-1159-0}{Few Body Syst. \textbf{57}, no.12, 1185-1212 (2016).}

%\cite{Chen:2016spr}
\bibitem{Chen:2016spr}
H.~X.~Chen, W.~Chen, X.~Liu, Y.~R.~Liu and S.~L.~Zhu,
A review of the open charm and open bottom systems,
\href{https://doi.org/10.1088/1361-6633/aa6420}{Rept. Prog. Phys. \textbf{80}, no.7, 076201 (2017).}

%\cite{Lebed:2016hpi}
\bibitem{Lebed:2016hpi}
R.~F.~Lebed, R.~E.~Mitchell and E.~S.~Swanson,
Heavy-Quark QCD Exotica,
\href{https://doi.org/10.1016/j.ppnp.2016.11.003}{Prog. Part. Nucl. Phys. \textbf{93}, 143-194 (2017).}

%\cite{Esposito:2016noz}
\bibitem{Esposito:2016noz}
A.~Esposito, A.~Pilloni and A.~D.~Polosa,
Multiquark Resonances,
\href{https://doi.org/10.1016/j.physrep.2016.11.002}{Phys. Rept. \textbf{668}, 1-97 (2017).}

%\cite{Guo:2017jvc}
\bibitem{Guo:2017jvc}
F.~K.~Guo, C.~Hanhart, U.~G.~Mei\ss{}ner, Q.~Wang, Q.~Zhao and B.~S.~Zou,
Hadronic molecules,
\href{https://doi.org/10.1103/RevModPhys.90.015004}{Rev. Mod. Phys. \textbf{90} (2018) no.1, 015004}
\href{https://doi.org/10.1103/RevModPhys.94.029901}{[erratum: Rev. Mod. Phys. \textbf{94} (2022) no.2, 029901].}

%\cite{Ali:2017jda}
\bibitem{Ali:2017jda}
A.~Ali, J.~S.~Lange and S.~Stone,
Exotics: Heavy Pentaquarks and Tetraquarks,
\href{https://doi.org/10.1016/j.ppnp.2017.08.003}{Prog. Part. Nucl. Phys. \textbf{97}, 123-198 (2017).}

%\cite{Brambilla:2019esw}
\bibitem{Brambilla:2019esw}
N.~Brambilla, S.~Eidelman, C.~Hanhart, A.~Nefediev, C.~P.~Shen, C.~E.~Thomas, A.~Vairo and C.~Z.~Yuan,
The $XYZ$ states: experimental and theoretical status and perspectives,
\href{https://doi.org/10.1016/j.physrep.2020.05.001}{Phys. Rept. \textbf{873}, 1-154 (2020).}

%\cite{Liu:2019zoy}
\bibitem{Liu:2019zoy}
Y.~R.~Liu, H.~X.~Chen, W.~Chen, X.~Liu and S.~L.~Zhu,
Pentaquark and Tetraquark states,
\href{https://doi.org/10.1016/j.ppnp.2019.04.003}{Prog. Part. Nucl. Phys. \textbf{107}, 237-320 (2019).}

%\cite{Chen:2022asf}
\bibitem{Chen:2022asf}
H.~X.~Chen, W.~Chen, X.~Liu, Y.~R.~Liu and S.~L.~Zhu,
An updated review of the new hadron states,
\href{https://doi.org/10.1088/1361-6633/aca3b6}{Rept. Prog. Phys. \textbf{86}, no.2, 026201 (2023).}

    %\cite{CMS:2023owd}
    \bibitem{CMS:2023owd}
    A.~Hayrapetyan \textit{et al.} [CMS],
    New Structures in the $J/\psi J/\psi$ Mass Spectrum in Proton-Proton Collisions at $\sqrt{s}$=13\,\,TeV,
\href{https://journals.aps.org/prl/abstract/10.1103/PhysRevLett.132.111901}{Phys. Rev. Lett. \textbf{132}, 111901 (2024).}

    %\cite{LHCb:2020bwg}
    \bibitem{LHCb:2020bwg}
    R.~Aaij \textit{et al.} [LHCb],
    Observation of structure in the $J /\psi$-pair mass spectrum,
\href{https://linkinghub.elsevier.com/retrieve/pii/S2095927320305685}{Sci. Bull. \textbf{65}, 1983-1993 (2020).}

    %\cite{ATLAS:2023bft}
    \bibitem{ATLAS:2023bft}
    G.~Aad \textit{et al.} [ATLAS],
    Observation of an Excess of Dicharmonium Events in the Four-Muon Final State with the ATLAS Detector,
\href{https://journals.aps.org/prl/abstract/10.1103/PhysRevLett.131.151902}{Phys. Rev. Lett. \textbf{131}, 151902 (2023).}


%\cite{liu:2020eha}
\bibitem{liu:2020eha}
M.~S.~liu, F.~X.~Liu, X.~H.~Zhong and Q.~Zhao,
Fully heavy tetraquark states and their evidences in LHC observations,
\href{https://journals.aps.org/prd/abstract/10.1103/PhysRevD.109.076017}{Phys. Rev. D \textbf{109}, 076017 (2024).}

%\cite{Tiwari:2021tmz}
\bibitem{Tiwari:2021tmz}
R.~Tiwari, D.~P.~Rathaud and A.~K.~Rai, Spectroscopy of all charm tetraquark states,
\href{https://link.springer.com/article/10.1007/s12648-022-02427-8}{
Indian J. Phys. \textbf{97}, 943-954 (2023).}

%\cite{Lu:2020cns}
\bibitem{Lu:2020cns}
Q.~F.~L\"u, D.~Y.~Chen and Y.~B.~Dong,
Masses of fully heavy tetraquarks $QQ {\bar{Q}} {\bar{Q}}$ in an extended relativized quark model,
\href{https://link.springer.com/article/10.1140/epjc/s10052-020-08454-1}{Eur. Phys. J. C \textbf{80}, 871 (2020).}

%\cite{Faustov:2020qfm}
\bibitem{Faustov:2020qfm}
R.~N.~Faustov, V.~O.~Galkin and E.~M.~Savchenko,
Masses of the $QQ\bar Q\bar Q$ tetraquarks in the relativistic diquark--antidiquark picture,
\href{https://journals.aps.org/prd/abstract/10.1103/PhysRevD.102.114030}{Phys. Rev. D \textbf{102}, 114030 (2020).}

%\cite{Zhang:2020xtb}
\bibitem{Zhang:2020xtb}
J.~R.~Zhang,
$0^{+}$ fully-charmed tetraquark states,
\href{https://journals.aps.org/prd/abstract/10.1103/PhysRevD.103.014018}{Phys. Rev. D \textbf{103}, 014018 (2021).}

%\cite{Li:2021ygk}
\bibitem{Li:2021ygk}
Q.~Li, C.~H.~Chang, G.~L.~Wang and T.~Wang,
Mass spectra and wave functions of $T_{QQ\bar{Q}\bar{Q}}$ tetraquarks,
\href{https://journals.aps.org/prd/abstract/10.1103/PhysRevD.104.014018}{Phys. Rev. D \textbf{104}, 014018 (2021).}

%\cite{Liu:2021rtn}
\bibitem{Liu:2021rtn}
F.~X.~Liu, M.~S.~Liu, X.~H.~Zhong and Q.~Zhao,
Higher mass spectra of the fully-charmed and fully-bottom tetraquarks,
\href{https://journals.aps.org/prd/abstract/10.1103/PhysRevD.104.116029}{Phys. Rev. D \textbf{104}, 116029 (2021).}

%\cite{Giron:2020wpx}
\bibitem{Giron:2020wpx}
J.~F.~Giron and R.~F.~Lebed,
Simple spectrum of $c\bar c c\bar c$ states in the dynamical diquark model,
\href{https://journals.aps.org/prd/abstract/10.1103/PhysRevD.102.074003}{Phys. Rev. D \textbf{102}, 074003 (2020).}

%\cite{Karliner:2020dta}
\bibitem{Karliner:2020dta}
M.~Karliner and J.~L.~Rosner,
Interpretation of structure in the di- $J/\psi$ spectrum,
\href{https://journals.aps.org/prd/abstract/10.1103/PhysRevD.102.114039}{Phys. Rev. D \textbf{102}, 114039 (2020).}

%\cite{Wang:2020ols}
\bibitem{Wang:2020ols}
Z.~G.~Wang,
Tetraquark candidates in the LHCb's di-$J/\psi$ mass spectrum,
\href{https://iopscience.iop.org/article/10.1088/1674-1137/abb080}{Chin. Phys. C \textbf{44}, 113106 (2020).}

%\cite{Ke:2021iyh}
\bibitem{Ke:2021iyh}
H.~W.~Ke, X.~Han, X.~H.~Liu and Y.~L.~Shi,
Tetraquark state $X(6900)$ and the interaction between diquark and antidiquark,
\href{https://link.springer.com/article/10.1140/epjc/s10052-021-09229-y}{Eur. Phys. J. C \textbf{81}, 427 (2021).}

%\cite{Zhu:2020xni}
\bibitem{Zhu:2020xni}
R.~Zhu,
Fully-heavy tetraquark spectra and production at hadron colliders,
\href{https://www.sciencedirect.com/science/article/pii/S0550321321000900?via%3Dihub}{Nucl. Phys. B \textbf{966}, 115393 (2021).}

%\cite{Jin:2020jfc}
\bibitem{Jin:2020jfc}
X.~Jin, Y.~Xue, H.~Huang and J.~Ping,
Full-heavy tetraquarks in constituent quark models,
\href{https://link.springer.com/article/10.1140/epjc/s10052-020-08650-z}{Eur. Phys. J. C \textbf{80}, 1083 (2020).}

%\cite{Yang:2021hrb}
\bibitem{Yang:2021hrb}
G.~Yang, J.~Ping and J.~Segovia,
Exotic resonances of fully-heavy tetraquarks in a lattice-QCD insipired quark model,
\href{https://journals.aps.org/prd/abstract/10.1103/PhysRevD.104.014006}{Phys. Rev. D \textbf{104}, 014006 (2021).}

%\cite{Albuquerque:2020hio}
\bibitem{Albuquerque:2020hio}
R.~M.~Albuquerque, S.~Narison, A.~Rabemananjara, D.~Rabetiarivony and G.~Randriamanatrika,
Doubly-hidden scalar heavy molecules and tetraquarks states from QCD at NLO,
\href{https://journals.aps.org/prd/abstract/10.1103/PhysRevD.102.094001}{Phys. Rev. D \textbf{102}, 094001 (2020).}

%\cite{Albuquerque:2021erv}
\bibitem{Albuquerque:2021erv}
R.~M.~Albuquerque, S.~Narison, D.~Rabetiarivony and G.~Randriamanatrika,
Doubly hidden $0^{++}$ molecules and tetraquarks states from QCD at NLO,
\href{https://www.sciencedirect.com/science/article/pii/S2405601421000328?via%3Dihub}{Nucl. Part. Phys. Proc. \textbf{312-317}, 120-124 (2021).}

%\cite{Wu:2022qwd}
\bibitem{Wu:2022qwd}
R.~H.~Wu, Y.~S.~Zuo, C.~Y.~Wang, C.~Meng, Y.~Q.~Ma and K.~T.~Chao,
NLO results with operator mixing for fully heavy tetraquarks in QCD sum rules,
\href{https://link.springer.com/article/10.1007/JHEP11(2022)023}{JHEP \textbf{11}, 023 (2022).}

%\cite{Asadi:2021ids}
\bibitem{Asadi:2021ids}
Z.~Asadi and G.~R.~Boroun,
Masses of fully heavy tetraquark states from a four-quark static potential model,
\href{https://journals.aps.org/prd/abstract/10.1103/PhysRevD.105.014006}{Phys. Rev. D \textbf{105}, 014006 (2022).}

%\cite{Yang:2020wkh}
\bibitem{Yang:2020wkh}
B.~C.~Yang, L.~Tang and C.~F.~Qiao,
Scalar fully-heavy tetraquark states $QQ^\prime {\bar{Q}} \bar{Q^\prime }$ in QCD sum rules,
\href{https://link.springer.com/article/10.1140/epjc/s10052-021-09096-7}{Eur. Phys. J. C \textbf{81}, 324 (2021).}

 %\cite{Chen:2020xwe}
    \bibitem{Chen:2020xwe}
    H.~X.~Chen, W.~Chen, X.~Liu and S.~L.~Zhu,
    Strong decays of fully-charm tetraquarks into di-charmonia,
\href{https://linkinghub.elsevier.com/retrieve/pii/S2095927320305740}{Sci. Bull. \textbf{65}, 1994-2000 (2020).}

%\cite{Deng:2020iqw}
    \bibitem{Deng:2020iqw}
    C.~Deng, H.~Chen and J.~Ping,
    Towards the understanding of fully-heavy tetraquark states from various models,   \href{https://journals.aps.org/prd/abstract/10.1103/PhysRevD.103.014001}{Phys. Rev. D \textbf{103}, 014001 (2021).}

    %\cite{Gordillo:2020sgc}
    \bibitem{Gordillo:2020sgc}
    M.~C.~Gordillo, F.~De Soto and J.~Segovia,
    Diffusion Monte Carlo calculations of fully-heavy multiquark bound states,  \href{https://journals.aps.org/prd/abstract/10.1103/PhysRevD.102.114007}{Phys. Rev. D \textbf{102}, 114007 (2020).}

 %\cite{Zhao:2020zjh}
    \bibitem{Zhao:2020zjh}
    Z.~Zhao, K.~Xu, A.~Kaewsnod, X.~Liu, A.~Limphirat and Y.~Yan,
    Study of charmoniumlike and fully-charm tetraquark spectroscopy, \href{https://journals.aps.org/prd/abstract/10.1103/PhysRevD.103.116027}{Phys. Rev. D \textbf{103}, 116027 (2021).}

    %\cite{Faustov:2021hjs}
    \bibitem{Faustov:2021hjs}
    R.~N.~Faustov, V.~O.~Galkin and E.~M.~Savchenko,
    Heavy tetraquarks in the relativistic quark model,
\href{https://journals.aps.org/prd/abstract/10.1103/PhysRevD.103.116027}{Universe \textbf{7}, 94 (2021).}

   %\cite{Wang:2021kfv}
    \bibitem{Wang:2021kfv}
    G.~J.~Wang, L.~Meng, M.~Oka and S.~L.~Zhu,
    Higher fully charmed tetraquarks: Radial excitations and P-wave states,
\href{https://journals.aps.org/prd/abstract/10.1103/PhysRevD.104.036016}{Phys. Rev. D \textbf{104}, 036016 (2021).}









\iffalse
    %\cite{Chen:2020xwe}
    \bibitem{Chen:2020xwe}
    H.~X.~Chen, W.~Chen, X.~Liu and S.~L.~Zhu,
    Strong decays of fully-charm tetraquarks into di-charmonia,
    Sci. Bull. \textbf{65}, 1994-2000 (2020),
    %doi:10.1016/j.scib.2020.08.038
    [arXiv:2006.16027 [hep-ph]].
    %51 citations counted in INSPIRE as of 10 Jul 2022
     
    %\cite{Jin:2020jfc}
    \bibitem{Jin:2020jfc}
    X.~Jin, Y.~Xue, H.~Huang and J.~Ping,
    Full-heavy tetraquarks in constituent quark models,
    Eur. Phys. J. C \textbf{80}, no.11, 1083 (2020),
    %doi:10.1140/epjc/s10052-020-08650-z
    [arXiv:2006.13745 [hep-ph]].
    %48 citations counted in INSPIRE as of 10 Jul 2022
        
    %\cite{Lu:2020cns}
    \bibitem{Lu:2020cns}
    Q.~F.~L\"u, D.~Y.~Chen and Y.~B.~Dong,
    Masses of fully heavy tetraquarks $QQ {\bar{Q}} {\bar{Q}}$ in an extended relativized quark model,
    Eur. Phys. J. C \textbf{80}, no.9, 871 (2020),
    %doi:10.1140/epjc/s10052-020-08454-1
    [arXiv:2006.14445 [hep-ph]].
    %61 citations counted in INSPIRE as of 10 Jul 2022
                
    %\cite{Yang:2020rih}
    \bibitem{Yang:2020rih}
    G.~Yang, J.~Ping, L.~He and Q.~Wang,
    A potential model prediction of fully-heavy tetraquarks $QQ\bar{Q}\bar{Q}$ ($Q=c, b$),
    [arXiv:2006.13756 [hep-ph]].
    %3 citations counted in INSPIRE as of 29 Jul 2020
            
    %\cite{Deng:2020iqw}
    \bibitem{Deng:2020iqw}
    C.~Deng, H.~Chen and J.~Ping,
    Towards the understanding of fully-heavy tetraquark states from various models,
    Phys. Rev. D \textbf{103}, no.1, 014001 (2021),
    %doi:10.1103/PhysRevD.103.014001
    [arXiv:2003.05154 [hep-ph]].
    %46 citations counted in INSPIRE as of 10 Jul 2022
            
    %\cite{Wang:2020ols}
    \bibitem{Wang:2020ols}
    Z.~G.~Wang,
    Tetraquark candidates in the LHCb's di-$J/\psi$ mass spectrum,
    Chin. Phys. C \textbf{44}, 113106 (2020),
    %doi:10.1088/1674-1137/abb080
    [arXiv:2006.13028 [hep-ph]].
    %39 citations counted in INSPIRE as of 10 Jul 2022
    
        
    %\cite{Chen:2020lgj}
    \bibitem{Chen:2020lgj}
    X.~Chen,
    Fully-charm tetraquarks: $cc\bar{c}\bar{c}$,
    [arXiv:2001.06755 [hep-ph]].
    %4 citations counted in INSPIRE as of 29 Jul 2020
    
    
    %\cite{Albuquerque:2020hio}
    \bibitem{Albuquerque:2020hio}
    R.~M.~Albuquerque, S.~Narison, A.~Rabemananjara, D.~Rabetiarivony and G.~Randriamanatrika,
    Doubly-hidden scalar heavy molecules and tetraquarks states from QCD at NLO,
    Phys. Rev. D \textbf{102}, no.9, 094001 (2020),
    %doi:10.1103/PhysRevD.102.094001
    [arXiv:2008.01569 [hep-ph]].
    %43 citations counted in INSPIRE as of 10 Jul 2022
    
    
    %\cite{Sonnenschein:2020nwn}
    \bibitem{Sonnenschein:2020nwn}
    J.~Sonnenschein and D.~Weissman,
    Deciphering the recently discovered tetraquark candidates around 6.9 GeV,
    Eur. Phys. J. C \textbf{81}, no.1, 25 (2021),
    %doi:10.1140/epjc/s10052-020-08818-7
    [arXiv:2008.01095 [hep-ph]].
    %32 citations counted in INSPIRE as of 10 Jul 2022
    
    
    %\cite{Giron:2020wpx}
    \bibitem{Giron:2020wpx}
    J.~F.~Giron and R.~F.~Lebed,
    Simple spectrum of $c\bar c c\bar c$ states in the dynamical diquark model,
    Phys. Rev. D \textbf{102}, no.7, 074003 (2020),
    %doi:10.1103/PhysRevD.102.074003
    [arXiv:2008.01631 [hep-ph]].
    %55 citations counted in INSPIRE as of 10 Jul 2022
    
   
    %\cite{Richard:2020hdw}
    \bibitem{Richard:2020hdw}
    J.~M.~Richard,
    About the $J/\psi$ $J/\psi$ peak of LHCb: fully-charmed tetraquark?,
    Sci. Bull. \textbf{65}, 1954-1955 (2020),
    %doi:10.1016/j.scib.2020.08.020
    [arXiv:2008.01962 [hep-ph]].
    %30 citations counted in INSPIRE as of 10 Jul 2022
    
    
    %\cite{Becchi:2020uvq}
    \bibitem{Becchi:2020uvq}
    C.~Becchi, J.~Ferretti, A.~Giachino, L.~Maiani and E.~Santopinto,
    A study of $c c\bar{c}\bar{c}$ tetraquark decays in 4 muons and in $D^{(*)} \bar{D}^{(*)}$ at LHC,
    Phys. Lett. B \textbf{811}, 135952 (2020),
    %doi:10.1016/j.physletb.2020.135952
    [arXiv:2006.14388 [hep-ph]].
    %37 citations counted in INSPIRE as of 10 Jul 2022
    
    
    %\cite{liu:2020eha}
    \bibitem{liu:2020eha}
    M.~S.~liu, F.~X.~Liu, X.~H.~Zhong and Q.~Zhao,
    Fully heavy tetraquark states and their evidences in LHC observations,
    Phys. Rev. D \textbf{109}, no.7, 076017 (2024),
    %doi:10.1103/PhysRevD.109.076017
    [arXiv:2006.11952 [hep-ph]].
    %76 citations counted in INSPIRE as of 22 Jul 2024
    
    
    %\cite{Bedolla:2019zwg}
    \bibitem{Bedolla:2019zwg}
    M.~A.~Bedolla, J.~Ferretti, C.~D.~Roberts and E.~Santopinto,
    Spectrum of fully-heavy tetraquarks from a diquark+antidiquark perspective,
    Eur. Phys. J. C \textbf{80}, no.11, 1004 (2020),
    %doi:10.1140/epjc/s10052-020-08579-3
    [arXiv:1911.00960 [hep-ph]].
    %70 citations counted in INSPIRE as of 10 Jul 2022
    
    
    %\cite{Chao:2020dml}
    \bibitem{Chao:2020dml}
    K.~T.~Chao and S.~L.~Zhu,
    The possible tetraquark states $cc \bar c \bar c$ observed by the LHCb experiment,
    Sci. Bull. \textbf{65}, no.23, 1952-1953 (2020),
    %doi:10.1016/j.scib.2020.08.031
    [arXiv:2008.07670 [hep-ph]].
    %31 citations counted in INSPIRE as of 10 Jul 2022
    
    
    %\cite{Karliner:2020dta}
    \bibitem{Karliner:2020dta}
    M.~Karliner and J.~L.~Rosner,
    Interpretation of structure in the di- $J/\psi$ spectrum,
    Phys. Rev. D \textbf{102}, no.11, 114039 (2020),
    %doi:10.1103/PhysRevD.102.114039
    [arXiv:2009.04429 [hep-ph]].
    %47 citations counted in INSPIRE as of 10 Jul 2022
    
    
    
    %\cite{Faustov:2020qfm}
    \bibitem{Faustov:2020qfm}
    R.~N.~Faustov, V.~O.~Galkin and E.~M.~Savchenko,
    Masses of the $QQ\bar Q\bar Q$ tetraquarks in the relativistic diquark--antidiquark picture,
    Phys. Rev. D \textbf{102}, no.11, 114030 (2020),
    %doi:10.1103/PhysRevD.102.114030
    [arXiv:2009.13237 [hep-ph]].
    %27 citations counted in INSPIRE as of 10 Jul 2022
    
    
    %\cite{Gordillo:2020sgc}
    \bibitem{Gordillo:2020sgc}
    M.~C.~Gordillo, F.~De Soto and J.~Segovia,
    Diffusion Monte Carlo calculations of fully-heavy multiquark bound states,
    Phys. Rev. D \textbf{102}, no.11, 114007 (2020),
    %doi:10.1103/PhysRevD.102.114007
    [arXiv:2009.11889 [hep-ph]].
    %26 citations counted in INSPIRE as of 10 Jul 2022  
    
    %\cite{Weng:2020jao}
    \bibitem{Weng:2020jao}
    X.~Z.~Weng, X.~L.~Chen, W.~Z.~Deng and S.~L.~Zhu,
    Systematics of fully heavy tetraquarks,
    Phys. Rev. D \textbf{103}, no.3, 034001 (2021),
    %doi:10.1103/PhysRevD.103.034001
    [arXiv:2010.05163 [hep-ph]].
    %35 citations counted in INSPIRE as of 10 Jul 2022
    
    
    %\cite{Zhang:2020xtb}
    \bibitem{Zhang:2020xtb}
    J.~R.~Zhang,
    $0^{+}$ fully-charmed tetraquark states,
    Phys. Rev. D \textbf{103}, no.1, 014018 (2021),
    %doi:10.1103/PhysRevD.103.014018
    [arXiv:2010.07719 [hep-ph]].
    %27 citations counted in INSPIRE as of 10 Jul 2022
    
    
    %\cite{Yang:2020wkh}
    \bibitem{Yang:2020wkh}
    B.~C.~Yang, L.~Tang and C.~F.~Qiao,
    Scalar fully-heavy tetraquark states $QQ^\prime {\bar{Q}} \bar{Q^\prime }$ in QCD sum rules,
    Eur. Phys. J. C \textbf{81}, no.4, 324 (2021),
    %doi:10.1140/epjc/s10052-021-09096-7
    [arXiv:2012.04463 [hep-ph]].
    %21 citations counted in INSPIRE as of 10 Jul 2022
    
    
    
    %\cite{Zhao:2020zjh}
    \bibitem{Zhao:2020zjh}
    Z.~Zhao, K.~Xu, A.~Kaewsnod, X.~Liu, A.~Limphirat and Y.~Yan,
    Study of charmoniumlike and fully-charm tetraquark spectroscopy,
    Phys. Rev. D \textbf{103}, no.11, 116027 (2021),
    %doi:10.1103/PhysRevD.103.116027
    [arXiv:2012.15554 [hep-ph]].
    %20 citations counted in INSPIRE as of 10 Jul 2022
 
    
    %\cite{Faustov:2021hjs}
    \bibitem{Faustov:2021hjs}
    R.~N.~Faustov, V.~O.~Galkin and E.~M.~Savchenko,
    Heavy tetraquarks in the relativistic quark model,
    Universe \textbf{7}, no.4, 94 (2021),
    %doi:10.3390/universe7040094
    [arXiv:2103.01763 [hep-ph]].
    %33 citations counted in INSPIRE as of 10 Jul 2022
    
    %\cite{Ke:2021iyh}
    \bibitem{Ke:2021iyh}
    H.~W.~Ke, X.~Han, X.~H.~Liu and Y.~L.~Shi,
    Tetraquark state $X(6900)$ and the interaction between diquark and antidiquark,
    Eur. Phys. J. C \textbf{81}, no.5, 427 (2021),
    %doi:10.1140/epjc/s10052-021-09229-y
    [arXiv:2103.13140 [hep-ph]].
    %18 citations counted in INSPIRE as of 10 Jul 2022
    
    
    %\cite{Yang:2021hrb}
    \bibitem{Yang:2021hrb}
    G.~Yang, J.~Ping and J.~Segovia,
    Exotic resonances of fully-heavy tetraquarks in a lattice-QCD insipired quark model,
    Phys. Rev. D \textbf{104}, no.1, 014006 (2021),
    %doi:10.1103/PhysRevD.104.014006
    [arXiv:2104.08814 [hep-ph]].
    %13 citations counted in INSPIRE as of 10 Jul 2022
    
    
    %\cite{Li:2021ygk}
    \bibitem{Li:2021ygk}
    Q.~Li, C.~H.~Chang, G.~L.~Wang and T.~Wang,
    Mass spectra and wave functions of $T_{QQ\bar{Q}\bar{Q}}$ tetraquarks,
    Phys. Rev. D \textbf{104}, no.1, 014018 (2021),
    %doi:10.1103/PhysRevD.104.014018
    [arXiv:2104.12372 [hep-ph]].
    %14 citations counted in INSPIRE as of 10 Jul 2022
    
    
    %\cite{Asadi:2021ids}
    \bibitem{Asadi:2021ids}
    Z.~Asadi and G.~R.~Boroun,
    Masses of fully heavy tetraquark states from a four-quark static potential model,
    Phys. Rev. D \textbf{105}, no.1, 014006 (2022),
    %doi:10.1103/PhysRevD.105.014006
    [arXiv:2112.11028 [hep-ph]].
    %1 citations counted in INSPIRE as of 10 Jul 2022
    
    
    %\cite{Kuang:2022vdy}
    \bibitem{Kuang:2022vdy}
    Z.~Kuang, K.~Serafin, X.~Zhao and J.~P.~Vary,
    All-charm tetraquark in front form dynamics,
    Phys. Rev. D \textbf{105}, 094028 (2022),
    %doi:10.1103/PhysRevD.105.094028
    [arXiv:2201.06428 [hep-ph]].
    %1 citations counted in INSPIRE as of 10 Jul 2022

    
    %\cite{Wu:2022qwd}
    \bibitem{Wu:2022qwd}
    R.~H.~Wu, Y.~S.~Zuo, C.~Y.~Wang, C.~Meng, Y.~Q.~Ma and K.~T.~Chao,
    NLO results with operator mixing for fully heavy tetraquarks in QCD sum rules,
    JHEP \textbf{11}, 023 (2022),
    %doi:10.1007/JHEP11(2022)023
    [arXiv:2201.11714 [hep-ph]].
    %20 citations counted in INSPIRE as of 22 Jul 2024
    
    
    %\cite{Wang:2021kfv}
    \bibitem{Wang:2021kfv}
    G.~J.~Wang, L.~Meng, M.~Oka and S.~L.~Zhu,
    Higher fully charmed tetraquarks: Radial excitations and P-wave states,
    Phys. Rev. D \textbf{104}, no.3, 036016 (2021),
    %doi:10.1103/PhysRevD.104.036016
    [arXiv:2105.13109 [hep-ph]].
    %11 citations counted in INSPIRE as of 10 Jul 2022
    \fi
    
    %\cite{Wang:2020wrp}
    \bibitem{Wang:2020wrp}
    J.~Z.~Wang, D.~Y.~Chen, X.~Liu and T.~Matsuki,
    Producing fully charm structures in the $J/\psi$ -pair invariant mass spectrum,
    \href{https://journals.aps.org/prd/abstract/10.1103/PhysRevD.103.L071503}{Phys. Rev. D \textbf{103}, 071503 (2021).}

    %\cite{Wang:2020tpt}
    \bibitem{Wang:2020tpt}
    J.~Z.~Wang, X.~Liu and T.~Matsuki,
    Fully-heavy structures in the invariant mass spectrum of $J/\psi \psi(3686)$, $J/\psi \psi(3770)$, $\psi(3686) \psi(3686)$, and $J/\psi \Upsilon(1S)$ at hadron colliders,   \href{https://linkinghub.elsevier.com/retrieve/pii/S0370269321001490}{Phys. Lett. B \textbf{816}, 136209 (2021).}

    %\cite{Wang:2022jmb}
    \bibitem{Wang:2022jmb}
    J.~Z.~Wang and X.~Liu,
    Improved understanding of the peaking phenomenon existing in the new di-J/\ensuremath{\psi} invariant mass spectrum from the CMS Collaboration,
\href{https://journals.aps.org/prd/abstract/10.1103/PhysRevD.106.054015}{Phys. Rev. D \textbf{106}, 054015 (2022).}

    %\cite{Gong:2020bmg}
    \bibitem{Gong:2020bmg}
    C.~Gong, M.~C.~Du, Q.~Zhao, X.~H.~Zhong and B.~Zhou,
    Nature of X(6900) and its production mechanism at LHCb,
\href{https://www.sciencedirect.com/science/article/pii/S0370269321007346?via%3Dihub}{Phys. Lett. B \textbf{824}, 136794 (2022).}

    %\cite{Dong:2020nwy}
    \bibitem{Dong:2020nwy}
    X.~K.~Dong, V.~Baru, F.~K.~Guo, C.~Hanhart and A.~Nefediev,
    Coupled-Channel Interpretation of the LHCb Double-~$J/\psi$~Spectrum and Hints of a New State Near the~ $J/\psi J/\psi$~~Threshold,  \href{https://journals.aps.org/prl/abstract/10.1103/PhysRevLett.126.132001}{Phys. Rev. Lett. \textbf{126}, 132001 (2021),} \href{https://journals.aps.org/prl/abstract/10.1103/PhysRevLett.127.119901}{[erratum: Phys. Rev. Lett. \textbf{127}, 119901 (2021)].}

    %\cite{Dong:2021lkh}
    \bibitem{Dong:2021lkh}
    X.~K.~Dong, V.~Baru, F.~K.~Guo, C.~Hanhart, A.~Nefediev and B.~S.~Zou,
    Is the existence of a $J/\psi J/\psi$ bound state plausible?,
\href{https://linkinghub.elsevier.com/retrieve/pii/S2095927321006265}{Sci. Bull. \textbf{66}, 2462-2470 (2021).}

    %\cite{Liang:2021fzr}
    \bibitem{Liang:2021fzr}
    Z.~R.~Liang, X.~Y.~Wu and D.~L.~Yao,
    Hunting for states in the recent LHCb di-J/\ensuremath{\psi} invariant mass spectrum,
\href{https://journals.aps.org/prd/abstract/10.1103/PhysRevD.104.034034}{Phys. Rev. D \textbf{104}, 034034 (2021).}

%\cite{Lu:2023aal}
    \bibitem{Lu:2023aal}
    Y.~Lu, C.~Chen, G.~Y.~Qin and H.~Q.~Zheng,
    A discussion on the anomalous threshold enhancement of
    $J/\psi \to \psi(3770)$ couplings and $X(6900)$ peak,
\href{https://iopscience.iop.org/article/10.1088/1674-1137/ad2361}{Chin. Phys. C \textbf{48}, 041001 (2024).}

    %\cite{Guo:2020pvt}
    \bibitem{Guo:2020pvt}
    Z.~H.~Guo and J.~A.~Oller,
    Insights into the inner structures of the fully charmed tetraquark state $X(6900)$,
\href{https://journals.aps.org/prd/abstract/10.1103/PhysRevD.103.034024}{Phys. Rev. D \textbf{103}, 034024 (2021).}

    %\cite{Zhuang:2021pci}
    \bibitem{Zhuang:2021pci}
    Z.~Zhuang, Y.~Zhang, Y.~Ma and Q.~Wang,
    Lineshape of the compact fully heavy tetraquark,
\href{https://journals.aps.org/prd/abstract/10.1103/PhysRevD.105.054026}{Phys. Rev. D \textbf{105}, 054026 (2022).}

    %\cite{Lyu:2021qsh}
    \bibitem{Lyu:2021qsh}
    Y.~Lyu, H.~Tong, T.~Sugiura, S.~Aoki, T.~Doi, T.~Hatsuda, J.~Meng and T.~Miyamoto,
    Dibaryon with Highest Charm Number near Unitarity from Lattice QCD,
\href{https://journals.aps.org/prl/abstract/10.1103/PhysRevLett.127.072003}{Phys. Rev. Lett. \textbf{127}, 072003 (2021).}
    
    %\cite{Mathur:2022ovu}
    \bibitem{Mathur:2022ovu}
    N.~Mathur, M.~Padmanath and D.~Chakraborty,
    Strongly Bound Dibaryon with Maximal Beauty Flavor from Lattice QCD,
\href{https://journals.aps.org/prl/abstract/10.1103/PhysRevLett.130.111901}{Phys. Rev. Lett. \textbf{130}, 111901 (2023).}


    %\cite{Bodwin:1994jh}
\bibitem{Bodwin:1994jh}
G.~T.~Bodwin, E.~Braaten and G.~P.~Lepage,
Rigorous QCD analysis of inclusive annihilation and production of heavy quarkonium,
\href{https://journals.aps.org/prd/abstract/10.1103/PhysRevD.51.1125}{Phys. Rev. D \textbf{51}, 1125-1171 (1995)}
\href{https://journals.aps.org/prd/abstract/10.1103/PhysRevD.55.5853}{[erratum: Phys. Rev. D \textbf{55}, 5853 (1997)]}.
%doi:10.1103/PhysRevD.55.5853
%[arXiv:hep-ph/9407339 [hep-ph]].
%3067 citations counted in INSPIRE as of 07 Aug 2025

%\cite{ATLAS:2015zdw}
\bibitem{ATLAS:2015zdw}
G.~Aad \textit{et al.} [ATLAS],
Measurement of the differential cross-sections of prompt and non-prompt production of $J/\psi $ and $\psi (2\mathrm {S})$ in $pp$ collisions at $\sqrt{s} = 7$ and 8 TeV with the ATLAS detector,
\href{https://link.springer.com/article/10.1140/epjc/s10052-016-4050-8}{Eur. Phys. J. C \textbf{76}, no.5, 283 (2016).}
%doi:10.1140/epjc/s10052-016-4050-8
%[arXiv:1512.03657 [hep-ex]].
%108 citations counted in INSPIRE as of 07 Aug 2025

%\cite{ATLAS:2014ala}
\bibitem{ATLAS:2014ala}
G.~Aad \textit{et al.} [ATLAS],
Measurement of $\chi_{c1}$ and $\chi_{c2}$ production with $\sqrt{s}$ = 7 TeV $pp$ collisions at ATLAS,
\href{https://link.springer.com/article/10.1007/JHEP07(2014)154}{JHEP \textbf{07}, 154 (2014).}
%doi:10.1007/JHEP07(2014)154
%[arXiv:1404.7035 [hep-ex]].
%102 citations counted in INSPIRE as of 07 Aug 2025

%\cite{CMS:2012qwg}
\bibitem{CMS:2012qwg}
S.~Chatrchyan \textit{et al.} [CMS],
Measurement of the Relative Prompt Production Rate of $\chi_{c2}$ and $\chi_{c1}$ in $pp$ Collisions at $\sqrt{s}=7$ TeV,
\href{https://link.springer.com/article/10.1140/epjc/s10052-012-2251-3}{Eur. Phys. J. C \textbf{72}, 2251 (2012).}
%doi:10.1140/epjc/s10052-012-2251-3
%[arXiv:1210.0875 [hep-ex]].
%90 citations counted in INSPIRE as of 07 Aug 2025

%\cite{LHCb:2013ofo}
\bibitem{LHCb:2013ofo}
R.~Aaij \textit{et al.} [LHCb],
Measurement of the relative rate of prompt $\chi_{c0}$, $\chi_{c1}$ and $\chi_{c2}$ production at $\sqrt{s}=7$TeV,
\href{https://link.springer.com/article/10.1007/JHEP10(2013)115}{JHEP \textbf{10}, 115 (2013).}
%doi:10.1007/JHEP10(2013)115
%[arXiv:1307.4285 [hep-ex]].
%92 citations counted in INSPIRE as of 07 Aug 2025

%\cite{He:2024lrb}
\bibitem{He:2024lrb}
Z.~G.~He, X.~B.~Jin, B.~A.~Kniehl and R.~Li,
Next-to-leading-order relativistic and QCD corrections to prompt pair photoproduction at future colliders,
\href{https://doi.org/10.1088/1674-1137/ad408f}{Chin. Phys. C \textbf{48}, no.8, 083107 (2024).}
%doi:10.1088/1674-1137/ad408f
%[arXiv:2404.08945 [hep-ph]].
%2 citations counted in INSPIRE as of 07 Aug 2025


    %\cite{Cheng:1992xi}
    \bibitem{Cheng:1992xi}
    H.~Y.~Cheng, C.~Y.~Cheung, G.~L.~Lin, Y.~C.~Lin, T.~M.~Yan and H.~L.~Yu,
    Chiral Lagrangians for radiative decays of heavy hadrons,
\href{https://journals.aps.org/prd/abstract/10.1103/PhysRevD.47.1030}{Phys. Rev. D \textbf{47}, 1030-1042 (1993).}
    
    %\cite{Yan:1992gz}
    \bibitem{Yan:1992gz}
    T.~M.~Yan, H.~Y.~Cheng, C.~Y.~Cheung, G.~L.~Lin, Y.~C.~Lin and H.~L.~Yu,
    Heavy quark symmetry and chiral dynamics,
\href{https://journals.aps.org/prd/abstract/10.1103/PhysRevD.46.1148}{Phys. Rev. D \textbf{46}, 1148-1164 (1992),}
\href{https://journals.aps.org/prd/abstract/10.1103/PhysRevD.55.5851}{[erratum: Phys. Rev. D \textbf{55}, 5851 (1997)].}
    %doi:10.1103/PhysRevD.46.1148
    %783 citations counted in INSPIRE as of 22 Jul 2024
    
    %\cite{Wise:1992hn}
    \bibitem{Wise:1992hn}
    M.~B.~Wise,
    Chiral perturbation theory for hadrons containing a heavy quark,
\href{https://journals.aps.org/prd/abstract/10.1103/PhysRevD.45.R2188}{Phys. Rev. D \textbf{45}, R2188 (1992).}
    %doi:10.1103/PhysRevD.45.R2188
    %885 citations counted in INSPIRE as of 22 Jul 2024
    
    %\cite{Burdman:1992gh}
    \bibitem{Burdman:1992gh}
    G.~Burdman and J.~F.~Donoghue,
    Union of chiral and heavy quark symmetries,
\href{https://linkinghub.elsevier.com/retrieve/pii/037026939290068F}{Phys. Lett. B \textbf{280}, 287-291 (1992).}
    %doi:10.1016/0370-2693(92)90068-F
    %689 citations counted in INSPIRE as of 22 Jul 2024
    
    %\cite{Casalbuoni:1996pg}
    \bibitem{Casalbuoni:1996pg}
    R.~Casalbuoni, A.~Deandrea, N.~Di Bartolomeo, R.~Gatto, F.~Feruglio and G.~Nardulli,
    Phenomenology of heavy meson chiral Lagrangians,
\href{https://linkinghub.elsevier.com/retrieve/pii/S0370157396000270}{Phys. Rept. \textbf{281}, 145-238 (1997).}

    %\cite{Ding:2021igr}
    \bibitem{Ding:2021igr}
    Z.~M.~Ding, H.~Y.~Jiang, D.~Song and J.~He,
    Hidden and doubly heavy molecular states from interactions $D^{(*)}_{(s)}{{\bar{D}}}^{(*)}_{s}$/$B^{(*)}_{(s)}{{\bar{B}}}^{(*)}_{s}$ and ${D}^{(*)}_{(s)}D_{s}^{(*)}$/${B}^{(*)}_{(s)}B_{s}^{(*)}$,
\href{https://link.springer.com/article/10.1140/epjc/s10052-021-09534-6}{Eur. Phys. J. C \textbf{81}, 732 (2021).}

    %\cite{Falk:1992cx}
    \bibitem{Falk:1992cx}
    A.~F.~Falk and M.~E.~Luke,
    Strong decays of excited heavy mesons in chiral perturbation theory,
\href{https://linkinghub.elsevier.com/retrieve/pii/037026939290618E}{Phys. Lett. B \textbf{292}, 119-127 (1992).}
    
    %\cite{Isola:2003fh}
    \bibitem{Isola:2003fh}
    C.~Isola, M.~Ladisa, G.~Nardulli and P.~Santorelli,
    Charming penguins in $B \to K^* \pi, ~ K(\rho,\omega,\phi)$ decays,
\href{https://journals.aps.org/prd/abstract/10.1103/PhysRevD.68.114001}{Phys. Rev. D \textbf{68}, 114001 (2003).}
    
    %\cite{Liu:2009qhy}
    \bibitem{Liu:2009qhy}
    X.~Liu, Z.~G.~Luo, Y.~R.~Liu and S.~L.~Zhu,
    X(3872) and Other Possible Heavy Molecular States,
\href{https://link.springer.com/article/10.1140/epjc/s10052-009-1020-4}{Eur. Phys. J. C \textbf{61}, 411-428 (2009).}
    
    %\cite{Chen:2019asm}
    \bibitem{Chen:2019asm}
    R.~Chen, Z.~F.~Sun, X.~Liu and S.~L.~Zhu,
    Strong LHCb evidence supporting the existence of the hidden-charm molecular pentaquarks,
\href{https://journals.aps.org/prd/abstract/10.1103/PhysRevD.100.011502}{Phys. Rev. D \textbf{100}, 011502 (2019).}

    %\cite{Huang:2020kxf}
    \bibitem{Huang:2020kxf}
    Q.~Huang, C.~W.~Shen and J.~J.~Wu,
    Detecting the pure triangle sigularity effect through the $\psi(2S) \to p \bar{p} \eta / p \bar{p} \pi^0$ process,
    \href{doi:10.1103/PhysRevD.103.016014}{Phys. Rev. D \textbf{103}, no.1, 016014 (2021).}
    %[arXiv:2011.14590 [hep-ph]].
    %10 citations counted in INSPIRE as of 12 Feb 2026

    %\cite{Dong:2020tdw}
    \bibitem{Dong:2020tdw}
    X.~K.~Dong, X.~H.~Mo, P.~Wang and C.~Z.~Yuan,
    Hadronic cross section of $e^+e^-$ annihilation at bottomonium energy region,
    \href{https://iopscience.iop.org/article/10.1088/1674-1137/44/8/083001}{Chin. Phys. C \textbf{44}, 083001 (2020).}

    %\cite{CMS:2025ecr}
    \bibitem{CMS:2025ecr}
    A.~Hayrapetyan \textit{et al.} [CMS],
    Determination of the spin and parity of all-charm tetraquarks,
    \href{https://www.nature.com/articles/s41586-025-09711-7}{Nature \textbf{648}, 58-63 (2025)}.
    %doi:10.1038/s41586-025-09711-7
    %[arXiv:2506.07944 [hep-ex]].
    %2 citations counted in INSPIRE as of 07 Aug 2025

\end{thebibliography}
\end{document}